\documentclass[doublecol]{epl2}
\usepackage{amsmath}\usepackage{amssymb}\usepackage{bm}\usepackage{psfrag}

\title{Transport properties of an interacting quantum dot with a non-uniform magnetization}
\shorttitle{Transport of a quantum dot with a magnetization} 

\author{N. Sedlmayr\footnote{nicholas.sedlmayr@physik.uni-halle.de}\inst{} \and J. Berakdar\inst{}}
\shortauthor{N. Sedlmayr \etal}

\institute{
  \inst{} Institut f\"ur Physik, Martin-Luther Universit\"at Halle-Wittenberg, Heinrich-Damerow-Str.4
06120 Halle (Saale), Germany\\
}

\abstract{
We study  the influence of the  non-homogeneity of a magnetization field on the behaviour of interacting
electrons in a quantum dot. In particular we investigate the magnetotransport properties
when the dot is weakly coupled to two ferromagnetic leads.
We take into account the interactions in the quantum dot non-perturbatively.
For a magnetization which varies slowly on the scale of the
Fermi wave length, the non-homogeneity effect is described by a gauge potential that can be treated perturbatively.}

\newcommand{\bpsi}{\bar{\psi}}

\newcommand{\bq}{\mathbf{q}}
\newcommand{\bk}{\mathbf{k}}
\newcommand{\br}{\mathbf{r}}

\newcommand{\cint}{\int_{c_{tt'}}}
\newcommand{\kint}{(\int_{c_{t\tau}}+\int_{c_{\tau t'}})\upd t}
\newcommand{\hh}{\hat{H}}
\newcommand{\ha}{\hat{a}}

\DeclareMathOperator{\sgn}{sgn}

\DeclareMathOperator{\tr}{tr}

\begin{document}

\maketitle

\section{Introduction}

The role of magnetization and interaction in determining the electronic and transport properties of  quantum dots has already received considerable attention. In particular, many efforts have been devoted to the regime where the dot can be considered as a two-level system\cite{zhang,takahashi,braun,martinek,weymann1} or as a double island\cite{weymann2}.
 On the other hand, recent experimental activities indicate a delicate role of the magnetization on the
 magnetotransport properties of a metallic dot. E.g.,
 up to several hundred per cent   ballistic
magnetoresistance was measured  for Ni \cite{1,2,4},  Co \cite{3} and
 Fe nano-islands connected to ferromagnetic leads (cf. Refs\cite{ralph,J_phys_c_08} for an overview and further references).
 Such metallic, magnetic nanodots  can no longer be reduced to a two-level system\cite{kg,sedl}.

In addition the magnetization may well not be homogenous, in particular  when the exchange length is on the scale of the dot size. Hence, we inspect the scenario of a nonhomogeneously magnetized, interacting dot with a  mean level spacing, $\delta$, which is much smaller than all other relevant energy scales.  At suitably low temperatures such quantum dot systems display the Coulomb blockade effect\cite{pusglaz}. Specifically, we consider two non-collinear ferromagnetic leads coupled to the quantum dot and investigate the transport properties. Such a set-up is also suited to investigate scanning tunneling microscopy measurements on islands which have non-homogenous magnetizations \cite{stm}. As far as we are aware such a system had yet to be studied.

The quantum dot may be viewed as the domain wall region between the ferromagnetic leads. For low-density diluted magnetic semiconductor wires we studied a similar situation using a Luttinger-liquid and a renormalization group approach \cite{miguel} under the assumption that the carriers Fermi-wavelength  is larger than the dot region; the effect of which can then be modeled   as a point-like, spin-dependent scatterer. These conditions are not applicable to the metallic case, because of the much smaller Fermi wave length the influence of the details of the magnetization profile need to be considered in this case.

\section{The model}

We consider a quantum dot with a  spatially non-uniform magnetization $\mathbf{M}(\br)$, which couples to the electron spin density with a strength $J$. The dot is weakly coupled to the ferromagnetic leads that have  uniform magnetizations aligned in opposite directions.  Due to the weak coupling to the leads we consider, to lowest order in the coupling, the quantum dot as being closed. The Hamiltonian for the dot region in terms of the free and interacting contributions reads
\begin{eqnarray}
\hh_0=\int
\upd\br\ha_\alpha^\dagger(\br)\hat{\xi}(\br)\ha_\alpha(\br)
-J\int \upd\br
\ha^\dagger_\alpha\bm{\sigma}_{\alpha\beta}.\mathbf{M}(\br)\ha_\beta
\end{eqnarray}
and
\begin{eqnarray}\label{hi}
\hh_I&=&\frac{E_c}{2}\hat{N}^2\quad\textrm{with}
\quad\hat{N}=\sum_{k\alpha}\ha_{k\alpha}^\dagger\ha_{k\alpha}.
\end{eqnarray}
We used units in which $\hbar=1$ and $k_B=1$  throughout. Repeated indices are summed over. $\alpha^\dagger_{k\alpha}$ is the creation operator for an electron of spin $\alpha$ and with a quantum number $k$ which labels the states in the dot. $\hat{\xi}=\hat{\varepsilon}-\mu$ is the dispersion for the quantum dot measured from the chemical potential. This chemical potential includes a gate voltage applied to the quantum dot. Eq.~\eqref{hi} describes a simplified Coulomb interaction\cite{kaa,cbrev} in a quantum dot with a charging energy $E_c=e^2/2C$, $C$ is the self-capacitance of the dot.  A vector gauge
transformation is performed to simplify the magnetization term, allowing us to use perturbation theory on the resulting introduced potential. After this transformation we will have a Zeeman
splitting term and a spin-dependent spatially varying potential,
$\hat{U}_{\alpha\beta}(\br)$ \cite{kor,tatara,dugaev1,dugaev2}. Perturbation theory is valid provided the magnetization varies
slowly compared with the Fermi wavelength of the electrons, which is usually the case for a metallic dot.
The gauge transformation\cite{kor,tatara,dugaev1,dugaev2} is
\begin{eqnarray}\label{gauge}
\begin{pmatrix}
\ha^{\textrm{old}}_1(\br)\\
\ha^{\textrm{old}}_2(\br)\end{pmatrix}
=\mathbf{T}(\br)
\begin{pmatrix}\ha^{\textrm{new}}_1(\br)\\
\ha^{\textrm{new}}_2(\br)\end{pmatrix}
\end{eqnarray}
where the unitary matrix $\mathbf{T}(\br)$ is defined such that
\begin{eqnarray}
\mathbf{T}^\dagger(\br)\vec{\bm{\sigma}}(\br).\vec{n}(\br)\mathbf{T}(\br)
&=&\bm{\sigma}^z\textrm{ where}\\
\vec{M}(\br)&=&\vec{n}(\br)M.
\end{eqnarray}
$\vec{n}(\br)$ is a unit vector and $M$ is the, spatially invariant, size of the magnetization.
After the transformation our new Hamiltonian can be written as
\begin{eqnarray}
\hh=\int
\upd\br\ha_\alpha^\dagger(\br)[\hat{\xi}\delta_{\alpha\beta}
-JM\sigma^z_{\alpha\beta}+\hat{U}_{\alpha\beta}(\br)]\ha_\beta(\br)+\hh_I
\end{eqnarray}
with the potential given by
\begin{eqnarray}\label{eq:u}
\hat{\mathbf{U}}(\br)=-\frac{1}{2m}[2\vec{\mathbf{A}}(\br).\partial_{\br}+
\partial_\br.\vec{\mathbf{A}}+\vec{\mathbf{A}}^2(\br)].
\end{eqnarray}
The vector potential is defined, in terms of the transformation, as $\vec{\mathbf{A}}(\br)=\mathbf{T}^\dagger(\br)\partial_{\vec{\br}}
\mathbf{T}(\br)$.

For the case in which the magnetization is translationally
invariant in the $x$ and $y$ plane we set $\vec{n}(\br)\to
\vec{n}(z)$ and parameterize in terms of an angular function  $\varphi(z)$:
\begin{eqnarray}
\vec{n}(z)=\begin{pmatrix}\sin[\varphi(z)]\\0\\ \cos[\varphi(z)]
\end{pmatrix}.
\end{eqnarray}
Micromagnetic simulations for a magnetic stripe  of a length $L$ and
 a width and a thickness  smaller than the exchange length deliver
 the angular profile   $\varphi(z)=\pi-\cos^{-1}[\tanh(z/L)]$ (cf. Ref.\cite{thia} and references
therein for further details and experiments).
 Thus eq.~(\ref{eq:u}) takes on the form
\begin{eqnarray}
\hat{\mathbf{U}}(z)=\mathbb{I}\frac{[\varphi'(z)]^2}{8m}+i\bm{\sigma}^y
\bigg[\frac{\varphi''(z)}{4m}+\frac{\varphi'(z)\partial_z}{2m}\bigg].
\end{eqnarray}

Firstly we must manipulate the interacting term so that our Hamiltonian is
quadratic instead of quartic. Working in the functional integral representation for the Green's
function with the above Hamiltonian\cite{negor}, in the Keldysh representation\cite{keldysh,rams}, we have
\begin{eqnarray}
iG_{kk'}(t,t')=\frac{1}{\mathcal{Z}}
T_{kq}T^\dagger_{q'k'}
\int \textrm{D}\psi \textrm{D}\bpsi\psi_{k}(t)\bpsi_{k'}(t')
e^{iS_c+iS_U}
\end{eqnarray}
with the action given by
\begin{eqnarray}
iS_c&=&\underbrace{i\sum_{k\alpha}\int_c\upd t\bpsi_{k\alpha}(t)[i\partial_t-\xi_k
+JM\sigma^z_{\alpha\alpha}]\psi_{k\alpha}(t)}_{\equiv iS_0}\nonumber\\&&
-\frac{E_c}{2}\int_c\upd t\big(\sum_{k\alpha}\bpsi_{k\alpha}(t)\psi_{k\alpha}(t)
\big)^2\textrm{ and}\\
iS_U&=&-i\sum_{\substack{kk'\\ \alpha\alpha'}}\int_c\upd t\bpsi_{k\alpha}(t)U_{\alpha\alpha'}
(k,k')\psi_{k'\alpha'}(t).
\end{eqnarray}
We define $\xi_{k\alpha}=\varepsilon_{k\alpha}-\mu=\varepsilon_k-\mu-JM\sigma^z_{\alpha\alpha}$ and $\bk\equiv\{k,\sigma\}$. The contour $c$ is the un-rotated Keldysh contour\cite{sedl} and all times are defined \emph{along} this contour.

We can rewrite the effect of the interaction using a Hubbard-Stratonovich transformation\cite{sedl}, introducing a bosonic field $\theta$:
\begin{eqnarray}\label{gav}
iG_{\bk\bk'}(t)\approx e^{-\frac{E_ct^2}{2\beta}-\frac{iE_ct\sgn(t)}{2}}
\langle \mathcal{Z}(\theta)iG^U_{\bk\bk'}(t)\rangle_\theta.
\end{eqnarray}
$\beta$ is the inverse temperature. Now
\begin{eqnarray}
iG^U_{\bk\bk'}(t,t')&=&\frac{1}{\mathcal{Z}}
T_{\bk\bq}T^\dagger_{\bq'\bk'}
\int \textrm{D}\psi \textrm{D}\bpsi\psi_{\bk}(t)\bpsi_{\bk'}(t')
e^{iS_\theta+iS_U}\nonumber\\
iS_\theta&=&i\sum_{\bk}\int_c\upd t\bpsi_{\bk}(t)[i\partial_t-\xi_\bk
+\theta/\beta]\psi_{\bk}(t)\nonumber\\
\mathcal{Z}(\theta)&=&\int \textrm{D}\psi \textrm{D}\bpsi e^{iS_\theta}=\prod_{\bk}\big[
1+e^{\theta-\beta\xi_{\bk}}\big].
\end{eqnarray}
The average in eq.~\eqref{gav} is defined as
\begin{eqnarray}
\langle\ldots\rangle_\theta=\frac{\int d\theta e^{\frac{\theta^2}{2E_c\beta}}\ldots}{
\int d\theta e^{\frac{\theta^2}{2E_c\beta}+
\ln[\mathcal{Z}_U(\theta)]}}.
\end{eqnarray}
This is our starting point for perturbation theory. After expanding in
$\mathbf{U}$ we can perform the averaging over $\theta$, this is achieved by
transforming the grand canonical ensemble into a sum over canonical ensembles.

Let us first define the contour
\begin{eqnarray}
\cint \upd t=\begin{cases}\int^t_{t'}\upd t&\text{if $t>t'$ on $c$ and}\\
\int_c\upd t-\int_t^{t'}\upd t&\text{if $t<t'$ on $c$.}\end{cases}
\end{eqnarray}
In standard diagrammatic perturbation theory\cite{agd} we have, to first order:
\begin{widetext}
\begin{eqnarray}\label{zgu}
\mathcal{Z}(\theta)iG^U_{\bk\bk'}(t,t')&\approx&\sum_{\bk\bk'}
T_{\bk\bq}T^\dagger_{\bq'\bk'} \big[\sgn(t-t')e^{-i\xi_{\bk}\cint
\upd t}\mathcal{Z}_{\bk}(\theta) e^{i\frac{\theta}{\beta}\cint \upd t}
\delta_{\bk\bk'}\nonumber\\&&-i\int_c\upd\tau
\sgn(t-\tau)\sgn(\tau-t')e^{-i\xi_{\bk}\int_{c_{t\tau}} \upd t
-i\xi_{\bk'}\int_{c_{\tau t'}} \upd t}
\mathcal{Z}_{\bk
\bk'}(\theta)e^{i\frac{\theta}{\beta}\kint}
U_{\bk\bk'}\big].\qquad{}
\end{eqnarray}
\end{widetext}
\begin{center} see eq.~\eqref{zgu}.\end{center}
Where we have additionally defined
\begin{eqnarray}
\mathcal{Z}_{\bk}(\theta)&=&\frac{\mathcal{Z}(\theta)}{1+
e^{-\beta\xi_{\bk}+\theta}}=
\prod_{\mathbf{n}\neq \bk}\big[1+
e^{-\beta\xi_{\mathbf{n}}+\theta}\big],\\
\mathcal{Z}_{\bk\bk'}(\theta)&=&\frac{\mathcal{Z}(\theta)}{(1+
e^{-\beta\xi_{\bk}+\theta})(1+
e^{-\beta\xi_{\bk'}+\theta})}\nonumber\\&&=
\prod_{\mathbf{n}\neq\{\bk,\bk'\}}\big[1+
e^{-\beta\xi_{\mathbf{n}}+\theta}\big].
\end{eqnarray}

We need to calculate terms such as
\begin{eqnarray}\label{thi}
\langle\mathcal{Z}_{\bk\bk'}(\theta)
e^{i\frac{\theta}{\beta}\kint}\rangle_\theta.
\end{eqnarray}
We transform to the sum over canonical
ensembles thusly:
\begin{eqnarray}
\mathcal{Z}_{\bk\bk'}(\theta)=\sum_{N=0}^\infty
Z_N(\varepsilon_\bk,\varepsilon_{\bk'})
e^{(\beta\mu+\theta)N}
\end{eqnarray}
with
\begin{eqnarray}
Z_N(\varepsilon_\bk,\varepsilon_{\bk'})=\oint\frac{\upd\varphi}{2\pi}
e^{-iN\varphi}\prod_{\mathbf{n}\neq \{\bk,\bk'\}} \big[1+e^{-\beta\varepsilon_\mathbf{n}+i\varphi}\big].
\end{eqnarray}
(We have introduced $E_N=E_cN^2/2-\mu N$.)
The function $Z_N(\varepsilon_\bk,\varepsilon_{\bk'})$ can be rewritten as
\begin{eqnarray}
\frac{Z_N(\varepsilon_\bk,\varepsilon_{\bk'})}{Z_N}=[1-F_N(\varepsilon_\bk)]
[1-F_N(\varepsilon_{\bk'})].
\end{eqnarray}
$F_N(\varepsilon_\bk)$ is the canonical $N$-particle distribution
function for being in any $N$-particle state containing the level
$\varepsilon_\bk$. For large $N$ we find $F_N(\varepsilon_\bk)\approx
f(\varepsilon_\bk)$, the Fermi distribution.
Combining all of the above results thus far we can write the Green's function to first order:
\begin{widetext}
\begin{eqnarray}
iG_{\bk\bk'}(t,t')&=&
T_{\bk\bq}T^\dagger_{\bq'\bk'}
i\tilde{G}^c_{\bq\bq'}(t,t')
-\frac{i}{Z}\bigg[
T_{\bk\bq}T^\dagger_{\bq'\bk'}U_{\bq\bq'}
[1-f(\varepsilon_{\bk})][1-f(\varepsilon_{\bk'})]
e^{-\frac{E_c(t-t')^2}{2\beta}-\frac{iE_c(t-t')\sgn(t-t')}{2}}
\nonumber\\ &&\times
\int_c\upd\tau\sgn(t-\tau)\sgn(\tau-t')
e^{-i\xi_{\bk}\int_{c_{t\tau}} \upd t
-i\xi_{\bk'}\int_{c_{\tau t'}} \upd t}
\sum_{N=0}^\infty e^{-\beta E_N}
e^{-iE_cN\int_Kdt+\frac{E_c}{2\beta}(\int_Kdt)^2}\bigg].\label{longeq}
\end{eqnarray}
\end{widetext}
\begin{center} see eq.~\eqref{longeq}.\end{center}
$i\tilde{G}^c_{k\gamma}(t,t')$ is the Coulomb blockade result in
the presence of a Zeeman splitting term\cite{sedl} and
\begin{eqnarray}
Z=\sum_{N=0}^\infty e^{-\beta E_N}.
\end{eqnarray}
The second order terms are calculated in the same manner.

\section{The density of states}

The standard formula for the density of states in terms of the advanced and retarded Green's functions is given by
\begin{equation}
\nu_\gamma(\omega)=\frac{1}{2\pi}\tr_{k}[iG^R_{k\gamma
k'\gamma}(\omega)-iG^A_{k\gamma k'\gamma}(\omega)].
\end{equation}
We find no  first order correction in $U$ to the standard Coulomb blockade result\cite{pusglaz}, only second order terms.
For small distances compared with the size of the dot L, $|z|\ll L$,
the variation of the magnetization  inside the dot is approximately
$\varphi(z)\approx\pi/2+z/L$. This approximation still remains very accurate up to $|z|\sim L$ and allows us to perform the necessary sums and integrals. Hence we use
\begin{eqnarray}
\mathbf{U}(k,k')&\approx&\bigg[\frac{\mathbb{I}}{8mL^2}+i{\bm\sigma}^y
\frac{ik}{2mL}\bigg]\delta_{kk'}\textrm{ and }\\
\mathbf{T}&\approx&\frac{1}{\sqrt{2}}\begin{pmatrix} 1 & -1 \\ 1 & 1
\end{pmatrix}\delta_{k-k',0}.
\end{eqnarray}
$m$ is the electron mass. The magnitude of the magnetization is taken to be such that $JM>E_c$.

\begin{figure}
\onefigure[width=\columnwidth]{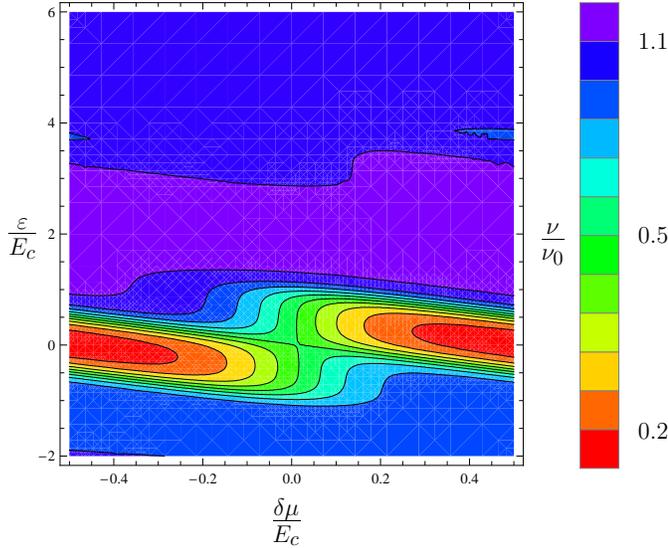}
\caption{A contour plot of the density of states $\nu(\varepsilon,\delta\mu)/\nu_0$ as a function of
 the energy ($\varepsilon/E_c$) and the gate voltage $\delta\mu/E_c$.}\label{fig1}
\end{figure}

\begin{figure}
\onefigure[width=\columnwidth]{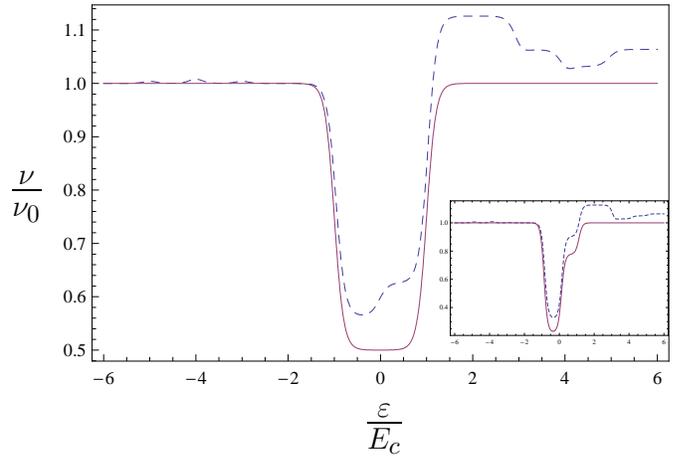}
\caption{The density of states $\nu(\varepsilon,\delta\mu)/\nu_0$ (dotted)
compared with the zeroth order form (full), as a function of energy $\varepsilon/E_c$. The energy, $\varepsilon$, is measured from the Fermi energy. The gate voltage is tuned here to the Coulomb blockade peak.  The inset shows a case intermediate between the Coulomb blockade valley and peak.}\label{fig2}
\end{figure}

\begin{figure}
\onefigure[width=\columnwidth]{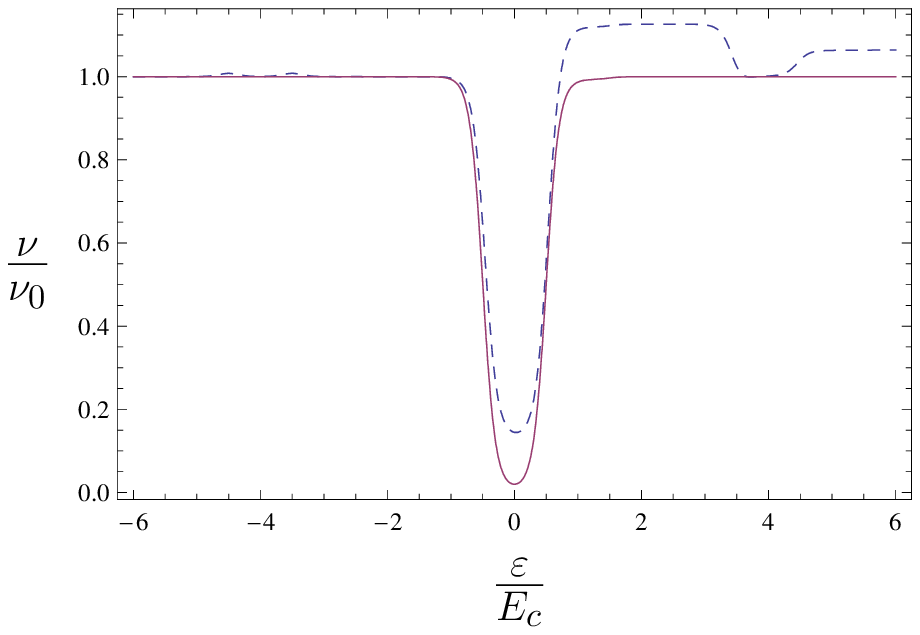}
\caption{The density of states $\nu(\varepsilon,\delta\mu)/\nu_0$ (dotted) compared with the zeroth order
form (full) as a function of energy $\varepsilon/E_c$. The energy, $\varepsilon$, is measured from the Fermi energy. The gate voltage is tuned in this case to the
centre of the Coulomb blockade valley.}\label{fig3}
\end{figure}

By varying an applied gate voltage to the dot we can shift the system through the Coulomb blockade valleys and peaks. We write the effect of the gate voltage as $\mu=E_c(N_0+\frac{1}{2})+\delta\mu$. $N_0$
is some (large) integer and $\delta\mu$ measures the distance from
the degeneracy point, i.e.~the peak in the Coulomb blockade regime. $\delta\mu=E_c/2$ is situated at the centre of the Coulomb valley. This
pattern repeats periodically in $E_c$.
The density of states is depicted in figure \ref{fig1}.  Figures
\ref{fig2} and \ref{fig3} show the corrected
density of states in comparison with the zeroth order density of
states for different values of the applied gate voltage. At this order in $U$ there is no difference between the density of states for spin up and for spin down.

We note that there is no longer any complete suppression of the density of states at any point.  The scattering opens up additional states on the scale of $JM$ around the Fermi level.
Below the Fermi energy we note the appearance of  subsidiary structures  in the density of states.  On closer inspection it becomes clear that their origin lies in the scattering from states at an energy $2JM$ below the Fermi level.

\section{Spin-dependent current}

For studying the transport properties we add two quasi-one
dimensional magnetic leads attached to the quantum dot by point contacts. The
quantum dot is described by the Hamiltonian
$\hat{H}=\hat{H}_0+\hat{H}_I+\hat{H}_U$ as before. We introduce a coupling
term, $\hat{H}_t$, between the dot and the leads. The current
through the dot is given by\cite{winmeir1,winmeir2}
\begin{eqnarray}
I&=&\dot{Q}=ei[\hat{H},\hat{N}]=ei[\hat{H}_t,\hat{N}]\\
\hat{H}&=&\hat{H}_{dot}+\hat{H}_t+\hat{H}_{leads}\\
\hat{H}_t&=&\sum_{\alpha \mathbf{n}\bk}[t_{\alpha
\mathbf{n}\bk} \hat{d}^{\dagger}_{\alpha \mathbf{n}}
\hat{a}_{\bk} +t^{*}_{\alpha
\mathbf{n}\bk}\hat{a}^{\dagger}_{\bk} \hat{d}_{\alpha
\mathbf{n}}].
\end{eqnarray}
Where $\alpha$ labels the leads (left and right),
$\hat{a}^\dagger$ is the creation operator for electrons in the
dot and $\hat{d}^\dagger_\alpha$ is the creation operator for
electrons in lead $\alpha$; the $t$'s describe tunneling between the dot and the leads. The current reads
\begin{eqnarray}
I&=&\frac{e}{2}\sum_{\alpha \bk\bk'}f_\alpha\int_{-\infty}^\infty
\frac{\upd\omega}{2\pi}[h^d(\omega)-h^\alpha(\omega)]\nonumber\\ && \times
[i\bar{G}^R_{\bk\bk'}(\omega)-i\bar{G}^A_{\bk\bk'}(\omega)]
\Gamma_{\bk'\bk}^\alpha(\omega),\\
\Gamma_{k'\sigma'k\sigma}^\alpha(\omega)&=&\sum_{\sigma''}
2\pi\nu^\alpha_{\sigma''} \langle
t_{\alpha\sigma'',k\sigma}t^*_{\alpha\sigma'',k'\sigma'}\rangle,
\end{eqnarray}
where we have assumed  $t$ to be independent of the lead states.
$i\bar{G}^{R/A}_{k\sigma k'\sigma'}(\omega)$ is the dot Green's
function coupled to the two leads. To lowest order in
$\nu^\alpha\Gamma^\alpha$ it is the unconnected dot Green's
function. $\nu^\alpha$ is the density of states for lead $\alpha$. The distribution function for lead $\alpha$ is $h^\alpha(\omega)=1-2f^\alpha(\omega)$ whereas
 $h^d(\omega)$ is the dot distribution function.
Demanding the flow of electrons in and out of the quantum dot to be balanced determines the steady-state lead distribution function and  the current is then
\begin{eqnarray}
I&=&\frac{e}{2}\int_{-\infty}^\infty\frac{\upd\omega}{2\pi}
[h^R(\omega)-h^L(\omega)]\frac{B_L(\omega)B_R(\omega)}{B_L(\omega)+B_R(\omega)},\nonumber\\
B_\alpha(\omega)&=&\sum_{k\sigma k'\sigma'}i\Delta G_{k\sigma
k'\sigma'}(\omega) \Gamma^\alpha_{k'\sigma'k\sigma}.
\label{B}\end{eqnarray}
We linearize the dispersion relation near the Fermi energy
and take $\Gamma$ to be a constant in $k$-space, equivalent to
assuming all momentum states in the dot are equally correlated. Furthermore
we assume that there is no spin scattering on tunneling and we have two ferromagnetic leads of opposite spin orientation. For simplicity we assume that $\Gamma^L=\Gamma^R$.

\begin{figure}
\onefigure[width=\columnwidth]{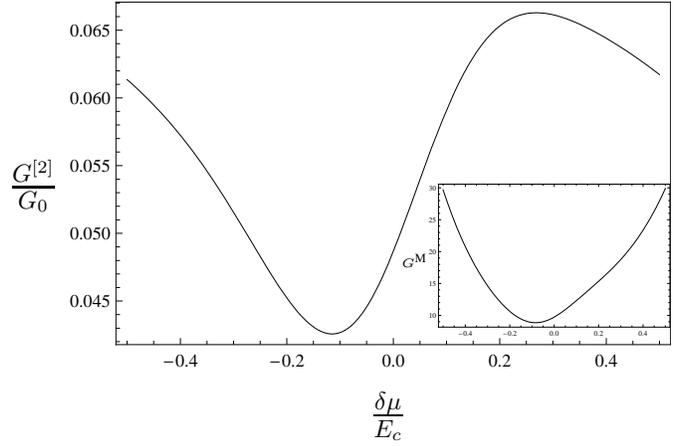}
\caption{The second order
correction to the linear differential conductance as a function of
$\delta\mu/E_c$. We plot $G^{[2]}/G_0$. The inset show the magnetotransport against $\delta\mu/E_c$. We define $G^{\textrm{M}}\equiv (G^{\uparrow\downarrow}-G^{\uparrow\uparrow})/G^{\uparrow\uparrow}$, $G^{\delta\gamma}$ being the differential conductance between leads with a spin direction of $\delta$ and $\gamma$.}\label{fig4}
\end{figure}

\begin{figure}
\onefigure[width=\columnwidth]{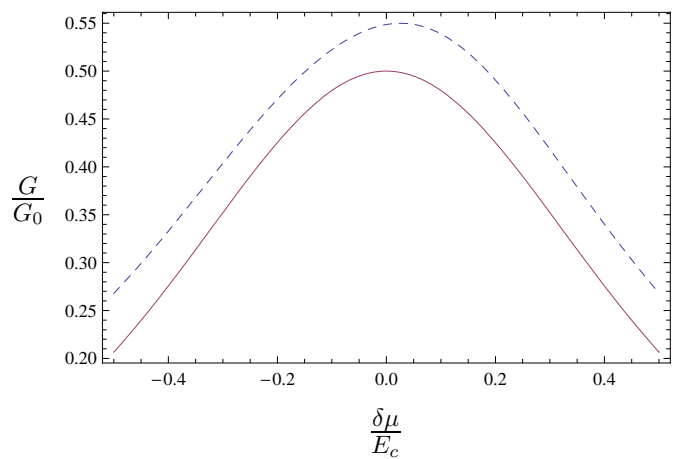}
\caption{The total linear
differential conductance (dotted) as a function of $\delta\mu/E_c$ compared
with the zeroth order linear differential
conductance (full). The differential conductance $G(\delta\mu)$ is scaled by $G_0=e^2\Gamma\nu_0$.}
\label{fig5}
\end{figure}

The linear differential conductance, for small biases, is given by
\begin{eqnarray}
G(\delta\mu)=\frac{dI}{dV}\bigg|_{V=V_L-V_R=0}.
\end{eqnarray}
The results are shown in figures \ref{fig4} and
\ref{fig5}. We plot the scaled differential conductance $G/G_0$ where $G_0=e^2\Gamma\nu_0$ is the conductance for temperatures much larger than the energy $E_c$ when charging effects inside the dot will play no role.
For a metallic system of size $L\approx 10^{-6}$m, $E_c\approx 20$meV. The temperature is taken to be $\approx 23$K, a tenth the size of the charging energy $E_c$. Also $E_c\approx U$.

As can be seen from figure \ref{fig5} the \emph{structure} of the second order correction to the linear differential conductance, see figure \ref{fig4}, is too small to be observed.   The main feature we note is that the Coulomb blockade valley is weakened, there is no longer a complete suppression of current at these points. The reason for this is clear if we look at the density of states in the valley, figure \ref{fig3}. The scattering from the potential $U$ introduces states into the previously empty region and hence it is always possible for electrons to tunnel through the dot. We emphasize that the difference between the curves in figure \ref{fig5} is \emph{absolute}. Thus it is possible to measure an absolute difference in the conductance due to the non-homogenous magnetization.  Additionally, although the structure of the second order correction is small, it can still be observed in the magneto-current. The inset of figure \ref{fig4} shows the magneto-current:
\begin{eqnarray}
G^{\textrm{M}}\equiv \frac{G^{\uparrow\downarrow}-G^{\uparrow\uparrow}}{G^{\uparrow\uparrow}}.
\end{eqnarray}
$G^{\uparrow\downarrow}$ is the current between non-collinear wires and $G^{\uparrow\uparrow}$ is the current between collinear wires, i.e.~the standard Coulomb blockade result with a homogenous magnetization in the dot.

\begin{figure}
\onefigure[width=\columnwidth]{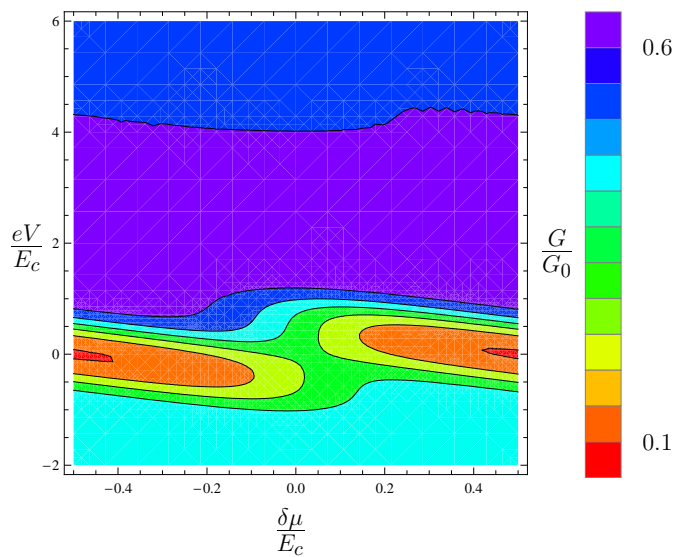}
\caption{The non-linear differential
conductance, $G/G_0$, as a function of $\delta\mu/E_c$ and
$eV/E_C$. The differential conductance $G(\delta\mu,eV)$ is scaled by $G_0=e^2\Gamma\nu_0$.}
\label{fig6}
\end{figure}

We will also consider the non-linear differential conductance. Let
us set $V_R=0$ and look at
\begin{eqnarray}
G(\delta\mu,V)=\frac{dI}{dV}\bigg|_{V_R=0}.
\end{eqnarray} This offers an approximate way of viewing the actual structure of the density of states\cite{aa}, which becomes exact at zero temperature. By tuning the gate voltage and the bias voltage we can map out the whole density of states, compare figures \ref{fig1} and \ref{fig6}. This pattern repeats periodically with the gate voltage, $\delta\mu$.

\section{Conclusions}

We have investigated the interplay of the Coulomb blockade and a non-homogenous magnetization for a quantum dot coupled to non-collinear ferromagnetic wires. We have shown that the scattering from the magnetization reduces the effect of the Coulomb blockade. By calculating the full profile for the density of states we clearly see the origin of this effect. States are created, by electrons scattering from the magnetization profile, in the previously depleted portion of the density of states. This occurs within an energy gap of the order of $JM$ from the Fermi level. Additionally we were able to calculate the full non-linear differential conductance, which would allow, in principle, the full mapping out of the density of states.

\acknowledgments
We thank V.~Dugaev for stimulating discussions. This research is supported by the DFG under SPP1165.


\begin{thebibliography}{0}

\bibitem{zhang}
\Name{Zhang P., Xue Q., Wang Y. \and Xie X.}
\REVIEW{Phys. Rev. Lett.}{89}{2002}{286803}

\bibitem{takahashi}
\Name{Takahashi S. \and Maekawa S.}
\REVIEW{Phys. Rev. Lett.}{80}{1998}{1758}

\bibitem{braun}
\Name{Braun M., Koenig J. \and Martinek J.}
\REVIEW{Phys. Rev. B}{70}{2004}{195345}

\bibitem{martinek}
\Name{Martinek J., Barnas J., Fert A. \and Maekawa S.}
\REVIEW{Journal of Applied Physics}{93}{2003}{8265}

\bibitem{weymann1}
\Name{Weymann I., Koenig J., Martinek J., Barnas J. \and Schoen G.}
\REVIEW{Phys. Rev. B}{72}{2005}{115334}

\bibitem{weymann2}
\Name{Weymann I. \and Barnas J.}
\REVIEW{Phys. Stat. Sol}{243}{2006}{239}

\bibitem{1}
\Name{Garcia, N., Munoz M. \and Zhao Y.-W.}
\REVIEW{Phys. Rev. Lett.}{82}{1999}{2923}

\bibitem{2}
\Name{Sullivan M. R., Boehm D. A., Ateya D. A., Hua S. Z. \and Chopra H. D.}
\REVIEW{Phys. Rev. B}{71}{2005}{024412}

\bibitem{4}
\Name{Garcia N., Munoz M., Qian G. G., Rohrer H., Saveliev I. G. \and Zhao Y.-W.}
\REVIEW{Appl. Phys. Lett.}{79}{2001}{4550}

\bibitem{3}
\Name{Chopra H. D., Sullivan M. R., Armstrong J. N. \and Hua S. Z.}
\REVIEW{Nature Mat.}{4}{2005}{832}

\bibitem{ralph}
\Name{Bolotin K. I., Kuemmeth F., Pasupathy A. N. \and Ralph D. C.}
\REVIEW{Nano Lett.}{6.1}{2006}{123}

\bibitem{J_phys_c_08}
\Name{Doudin B.,\and Viret M. J.}
\REVIEW{Phys. Condens. Matter}{20}{2008}{083201}

\bibitem{kg}
\Name{Kamenev. A. \and Gefen Y.}
\REVIEW{Phys. Rev. B}{54}{1996}{5428}

\bibitem{sedl}
\Name{Sedlmayr N., Yurkevich I. \and Lerner I.}
\REVIEW{Europhys. Lett.}{76}{2006}{109}

\bibitem{pusglaz}
\Name{Pustilnik M. \and Glazman L.}
\REVIEW{J. Phys.: Cond. Matter}{16}{2004}{5528}

\bibitem{stm}
\Name{Mironets O., {\it et al.}}
\REVIEW{Phys. Rev. Lett.}{100}{2008}{096103};
\Name{Pietzsch O., {\it et al.}}
\REVIEW{Phys. Rev. Lett.}{96}{2006}{237203}.

\bibitem{miguel}
\Name{Ara\' ujo M. A. N., Berakdar J., Dugaev V. K. \and Vieira V. R.}
\REVIEW{Phys. Rev. B}{76}{2007}{205107};
\REVIEW{Phys. Rev. B}{74}{2006}{224429};
\REVIEW {Physica E}{40}{2008}{1736}

\bibitem{kaa}
\Name{Kurland I., Aleiner I. \and Altshuler B.}
\REVIEW{Phys. Rev. B}{62}{2000}{14866}

\bibitem{cbrev}
\Name{Aleiner I., Brouwer P. \and Glazman L.}
\REVIEW{Phys. Rep.}{358}{2002}{309}

\bibitem{kor}
\Name{Korenman V., Murray J. L. \and Prange R. E.}
\REVIEW{Phys. Rev. B}{16}{1977}{4032}

\bibitem{tatara}
\Name{Tatara G. \and Fukuyama H.}
\REVIEW{Phys. Rev. Lett.}{78}{1997}{3773}

\bibitem{dugaev1}
\Name{Dugaev V., Barnas J., Lusakowski A. \and Turski L.}
\REVIEW{Phys. Rev. B}{65}{2002}{224419}

\bibitem{dugaev2}
\Name{Dugaev, V. K., Barnas, J. \and Berakdar, J.}
\REVIEW{Journal of Physics A}{36}{2003}{9263}

\bibitem{thia}
  \Name{Thiaville, A. \and  Nakatani, Y.}
  \Book{Spin Dynamics in Confined Magnetic Structures III, Topics in Appl. Physics
        \textbf{101}, 161–205 (2006), B. Hillebrands, A. Thiaville (Eds.):}
  \Publ{Springer-Verlag, Berlin}
  \Year{2006}.

\bibitem{negor}
  \Name{Negele J. \and Orland H.}
  \Book{Quantum Many-Particle Systems}
  \Publ{Addison-Wesley, Redwood City, California, Wokingham}
  \Year{1988}.

\bibitem{keldysh}
\Name{Keldysh L.}
\REVIEW{Zh. Eksp. Teor. Fiz.}{74}{1964}{1538}

\bibitem{rams}
\Name{Rammer J. \and Smith H.}
\REVIEW{Rev. Mod. Phys.}{58}{1986}{323}

\bibitem{agd}
  \Name{Abrikosov A., Gorkov L. \and Dzyaloshinski I.}
  \Book{Methods of Quantum Field Theory in Statistical Physics}
  \Publ{Dover, New York}
  \Year{1975}.

\bibitem{winmeir1}
\Name{Juaho A., Wingreen N. \and Meir Y.}
\REVIEW{Phys. Rev. B}{50}{1994}{5528}

\bibitem{winmeir2}
\Name{Wingreen N. \and Meir Y.}
\REVIEW{Phys. Rev. Lett.}{68}{1992}{2512}

\bibitem{aa}
  \Name{Altshuler B. \and Aronov A.}
  \Book{Electron-Electron Interactions in Disordered Conductors}
  \Publ{North-Holland, New York}
  \Year{1985}.


\end{thebibliography}
\end{document}